# Tether-cutting and Overlying Magnetic Reconnections in an MHD Simulation of Prominence-cavity System

Tie Liu[1,2] and Yingna Su[1,2]

[1]Key Laboratory for Dark Matter and Space Science, Purple Mountain Observatory, CAS, Nanjing 210008, China
[2]School of Astronomy and Space Science, University of Science and Technology of China, Hefei, Anhui 230026, China

## ABSTRACT

We investigate the magnetic reconnection in an MHD simulation of a coronal magnetic flux rope (MFR) confined by a helmet streamer, where a prominence-cavity system forms. This system includes a hot cavity surrounding a prominence with prominence horns and a central hot core above the prominence. The evolution of the system from quasi-equilibrium to eruption can be divided into four phases: quasi-static, slow rise, fast rise, and propagation phases. The emerged MFR initially stays quasi-static and magnetic reconnection occurs at the overlying high-Q (squashing factor) apex region, which gradually evolves into a hyperbolic flux tube (HFT). The decrease of the integrated magnetic tension force (above the location of the overlying reconnection) is due to the removal of overlying confinement by the enhanced overlying reconnection between the MFR and the overlying fields at the apex HFT, thus engines the slow rise of the MFR with a nearly constant velocity. Once the MFR reaches the regime of torus instability, another HFT immediately forms at the dip region under the MFR, followed by the explosive flare reconnection. The integrated resultant force (above the location of the flare reconnection) exponentially increases, which drives the exponential fast rise of the MFR. The system enters the propagation phase, once its apex reaches the height of about one solar radius above the photosphere. The simulation reproduces the main processes of one group of prominence eruptions especially those occurring on the quiet Sun.

## 1. INTRODUCTION

The most violently explosive phenomena in the solar corona are coronal mass ejections (CMEs) which eject a large amount of plasmas and magnetic fluxes to the solar-terrestrial space (Yashiro et al. 2004). Observations show that solar filament eruptions are often associated with CMEs (e.g., Webb & Hundhausen 1987; Gilbert et al. 2000; Gopalswamy 2003; Jing et al. 2004; McCauley et al. 2015). Some theoretic models involve magnetic flux rope (MFR) to explain the support of filament materials (e.g. Kuperus & Raadu 1974; Gibson et al. 2006). A large number of magnetohydrodynamics (MHD) simulations (e.g., Kliem et al. (2013) and Fan (2016)) have considered MFR to simulate the initiation of CMEs and obtain results similar to observations. Xia et al. (2014) and Xia & Keppens (2016) carry out the first three dimensional (3D) MHD simulation of a stable MFR with prominence formation, which shows fine scale and highly dynamic filaments materials and reproduces various features, such as the coronal cavity and filaments observed by SDO/AIA (Su et al. 2015). Fan (2017, 2018) and Fan & Liu (2019) has performed 3D MHD simulations of prominence-forming flux ropes to simulate CMEs associated with prominence eruptions.

It is well accepted that magnetic reconnection (MR) plays an important role in solar eruptions. Tether-cutting reconnection due to flux cancellations often plays an important role in the formation and eruption of filaments (van Ballegooijen & Martens 1989; Moore et al. 2001). In a single bipolar sheared field, flux cancellations at the magnetic polarity inversion line lead to magnetic reconnection that translates two sheared arcades into one sigmoidal flux rope,

ynsu@pmo.ac.cn

which then erupts. In the magnetic breakout model proposed by Antiochos et al. (1999), breakout reconnection at an overlying null point in a quadrupolar magnetic configuration can remove the confinement of the overlying flux, then allow the sheared core flux to erupt. Even in simulations where the loss of equilibrium of an MFR is the primary mechanism for CME initiation, magnetic reconnections are found to play an important role in the build-up and evolution of the flux rope leading up to the eruption and also in sustaining the acceleration of the flux rope (e.g. Aulanier et al. 2010; Fan 2010). Localized heating and dramatically changeable connection of the magnetic filed lines are characteristics of magnetic reconnection in MHD simulations. Normally the movement of the magnetic field line footpoints and the transformation of the magnetic field line geometrical properties are detected when magnetic reconnection occurs (Aulanier & Dudík 2019). In this work, we study magnetic reconnections mainly by tracing the localized heating and identifying the connection changes of the magnetic filed lines.

Observational analyses by Zhang et al. (2001, 2004) suggest that the kinematic evolution of CMEs can be divided into three phases: slow-rise phase, impulsive main-acceleration phase, and propagation phase with only slowly varying velocity. Filament/prominence eruptions are often characterized into slow rise and fast rise phases, based on a series of observational investigations (e.g., Sterling et al. 2007, 2011). Different kinematic characteristics indicate that different physical processes dominate in these phases.

Fan & Liu (2019) has carried out a 3D MHD simulation of prominence-cavity system evolving from quasi-equilibrium to eruption, and focused on the magnetic structure of the various observed features in the system such as the prominence "horns" and the central cavity. In this work, we examine the evolution of the prominence-cavity system and focus on the corresponding magnetic reconnections. In theory and simulations (e.g., ?Kliem et al. 2013), magnetic reconnection is usually driven by anomalous resistivity or artificial diffusion when magnetic field lines in opposite directions are close to each other. In the simulation understudy, magnetic reconnection is driven by numerical diffusion. A brief description of the MHD simulation is given in Section 2. The calculation method of the Q Factor is described in Section 3. In Section 4 we present an overview evolution of the prominence-cavity system, especially the magnetic flux rope and the corresponding filaments. We perform topological analysis and discuss the corresponding magnetic reconnection in Section 5. In Section 6 we present the summary and discussions.

## 2. MODEL DESCRIPTION

This paper investigates magnetic reconnection in an MHD simulation presented by Fan & Liu (2019), in which we focus on the magnetic structure of a prominence-cavity system which mainly contains four parts: the prominence, U-shape horns, hot core and cavity which correspond to the dips, shallow dips, axis and apex boundary of the magnetic flux rope. Detailed descriptions of the MHD simulation are presented in Section 2 of Fan & Liu (2019) and Sections 2 and 3.1 of Fan (2017), and a brief overview of the MHD simulation is given below.

We solve the semi-relativistic MHD equations (Gombosi et al. 2002; Rempel 2017) by the "Magnetic Flux Eruption" (MFE) code. The non-adiabatic effects such as an empirical coronal heating, the field-aligned electron heat conduction and optically thin radiative cooling are included in the energy equation, which contributes to the formation of prominence condensation in the emerged MFR. The simulation domain with a grid of $504(r) \times 196(\theta) \times 960(\varphi)$ is in spherical geometry ($r \in [R_\odot, 11.47 R_\odot]$, $\theta \in [75°, 105°]$, and $\varphi \in [-75°, 75°]$, where $R_\odot$ is the solar radius). The grids are uniform in $\theta$ and $\varphi$ directions. In the r direction, the grids are stretched with a grid size of $0.002727 R_\odot$ (i.e., $dr = 1.898$ Mm) for $r < 1.79 R_\odot$ and geometrically increase to about $0.19 R_\odot$ at the outer boundary ($r = 11.47 R_\odot$). The boundary conditions and initialization of hydrostatic atmosphere with a specified temperature profile are similar to those in Fan (2017, 2018). The emergence of a magnetic flux torus is imposed at the lower boundary of the initial streamer field with the speed of 1.95 km s$^{-1}$ as described in Fan (2017). The driving flux emergence is stopped at $t = 8.42$ hour when the total twist in the emerged flux rope reaches about 1.76 winds, then the prominence-cavity system enters the quasi-static phase. The main changes of the setup in the current simulation are the empirical coronal heating and field strength of the bipolar band. Two exponentially decaying components (Equation 1 in Fan & Liu (2019)) are included in the empirical coronal heating which contributes to open up the ambient magnetic field and promote the prominence condensation in the MFR. With the narrow bipolar band (the narrow-streamer solution in Fan 2017), we double the field strength and construct a prominence-cavity system (Figures 7 and 8 in Fan & Liu (2019)), whose cavity and prominence heights are lower than those in the case of Fan (2018), where the reproduced prominence and cavity are too high compared with the observations. As a result, we construct a prominence-cavity system which contains a long twisted MFR with ambient magnetic fields matching the typical observed values.

## 3. THE CALCULATION OF THE Q FACTOR

Priest & D´emoulin (1995) defines the location where the linkage of the magnetic field lines changes drastically as the Quasi-separatrix layers (QSLs). Magnetic reconnection takes place preferentially at the QSLs. Titov et al. (2002) proposes the squashing factor Q, which measures the deformation of a magnetic flux tube when mapping from one foot point to the other. The QSLs are defined as locations with high Q. The hyperbolic flux tubes (HFTs) defined as the intersections of two QSLs are locations with high Q factor and high incidence of magnetic reconnection. Good correspondence between magnetic topology and magnetic reconnections has been identified in previous studies such as Savcheva et al. (2015); Zhao et al. (2016); ?. In this study, we compute the Q factor to identify QSLs and investigate the corresponding magnetic reconnection in an MHD simulation of a prominence-cavity system.

The Q factor is calculated according to method 3 (i.e., equations 21 and 22) proposed by Pariat, E. & D´emoulin, P. (2012)). The plane r = 0 is chosen as the reference boundary. To calculate the Q factor at a point $r_c$ in the simulation domain, we first determine the magnetic field line passing through this point and find its two footpoints on the reference boundary. If $P_c$ represents the plane perpendicular to the magnetic field line at $r_c$, another four points on the plane $P_c$ with a fixed orthogonal distance δ from $r_c$ are found and four field lines are integrated from these four points. And then the corresponding footpoints of the four field lines on the reference boundary are identified. With the information of these footpoints on the reference boundary, we calculate the Q factor at point $r_c$ using the method 3 of Pariat, E. & D´emoulin, P. (2012) in which they point out that more and more accurate Q calculations can be obtained by using smaller and smaller values of the fixed distance δ. We set δ as 1/200, 000, 000 of the grid size on the reference boundary after several attempts. In our calculations, we define narrow sheet regions where the Q factor is greater than 100 as QSLs. HFTs are detected as the intersections of two QSLs.

## 4. OVERVIEW OF THE EVOLUTION OF THE PROMINENCE-FORMING CORONAL FLUX ROPE

Figure 1 shows the evolution of the prominence-cavity system from t = 8.42 hour to t = 20 hour during which a prominence eruption is identified. According to the height-time plot in Figure 1 as well as the velocity-time and acceleration-distance plots in Figures 3 and 4 of Fan & Liu (2019), we divide the evolution of the system into four phases, namely, the quasi-static (8.42 − 17 hour), slow rise (17 − 17.91 hour), fast rise/impulsive main acceleration (17.91 − 18.95 hour), and propagation (after 18.95 hour) phases. The division criteria are described as follows.

From 8.42 hour to 17 hour, the emerged flux rope stays in quasi-static state, during which the velocity remains nearly 0 (Figure 3 in Fan & Liu (2019)). Monotonic acceleration starts at 17 hour, which is defined as the beginning of the slow rise phase. In order to understand the evolution of eruptive prominences and CMEs observed at the low corona, Cheng et al. (2013) and Cheng et al. (2020) propose a fitting function $h(t) = c_0 e^{(t-t_0)/\tau} + c_1(t - t_0) + c_2$, in which a linear equation fitting the slow rise phase and an exponential equation fitting the fast rise phase. The onset of the fast rise phase is defined as a time when the exponential term starts to take over (equal) the linear term: $t_{onset} = \tau \ln(c_1\tau/c_0) + t_0$. This method is applied to fit the height-time plot of the MFR in the MHD simulation shown as red plus signs in Figure 1, the fitted curve is shown in blue and the five free parameters are also listed. The fitting starts at 17 hour, before which the rise velocity of the MFR is too small, and including a much earlier part of the slow rise is likely to obtain a much earlier onset time than reality (Cheng et al. 2020). According to this fitting, the onset time of the fast rise phase is identified as 17.91 hour. The fitting becomes worse after 18.95 hour when the apex of the axial field line of the emerged flux rope reaches the height of about one solar radius where the acceleration reduces to nearly 0 (Figures 4 of Fan & Liu 2019). Therefore, we define 18.95 hour as the beginning of the propagation phase which is characterized by constant or slowly decreasing speed.

We choose 156 magnetic field lines with fixed footpoints to represent the MFR. Panels a1-a4 of Figure 1 show snapshots of selected magnetic field lines colored by temperature at the quasi-static phase (at t = 15.86 hour), slow rise phase (at t = 17.84 hour) and fast rise phase (t = 18.63 hour and t = 18.83 hour). The corresponding snapshots of the density iso-surface are displayed in panels b1-b4, which indicates that the filament is located at the dips of magnetic field lines during the quasi-static phase, then partially erupts with the MFR (panels b2-b4).

Figure 2 shows snapshots of the temperature (row 1), density (row 2), current density normalized by field strength $J/B = |\nabla \times B|/B$ (row 3), and the computed squashing factor Q (row 4) in the central vertical slice of the MFR at time instances corresponding to the first three panels in Figure 1. As shown in Figure 2, we identify the following three distinct features at the quasi-static phase: (1) a hot core with relatively high temperature and density in the center of the MFR (marked with red arrow and "Hot core" in the first column). (2) A hot rim at the boundary of the low density cavity, with its apex marked with cyan arrow and "Apex" in the first column. The rim and its apex possess the highest temperature (panel a1), a low density (panel b1), and a high Q factor (panel d1) in the flux rope cross-section. (3) The dip region (marked with green arrow and "Dip") where the prominence resides has the lowest temperature (panel a1), a high density (panel b1), and a high Q factor (panel d1). The apex region corresponds to the top of the MFR where the overlying magnetic field lines connect very differently from the lower MFR, resulting in the sharp variety of the distributions of temperature, density, J/B and Q factor. Filaments are located at the dips of the MFR and the hot core corresponds to the axis of the MFR.

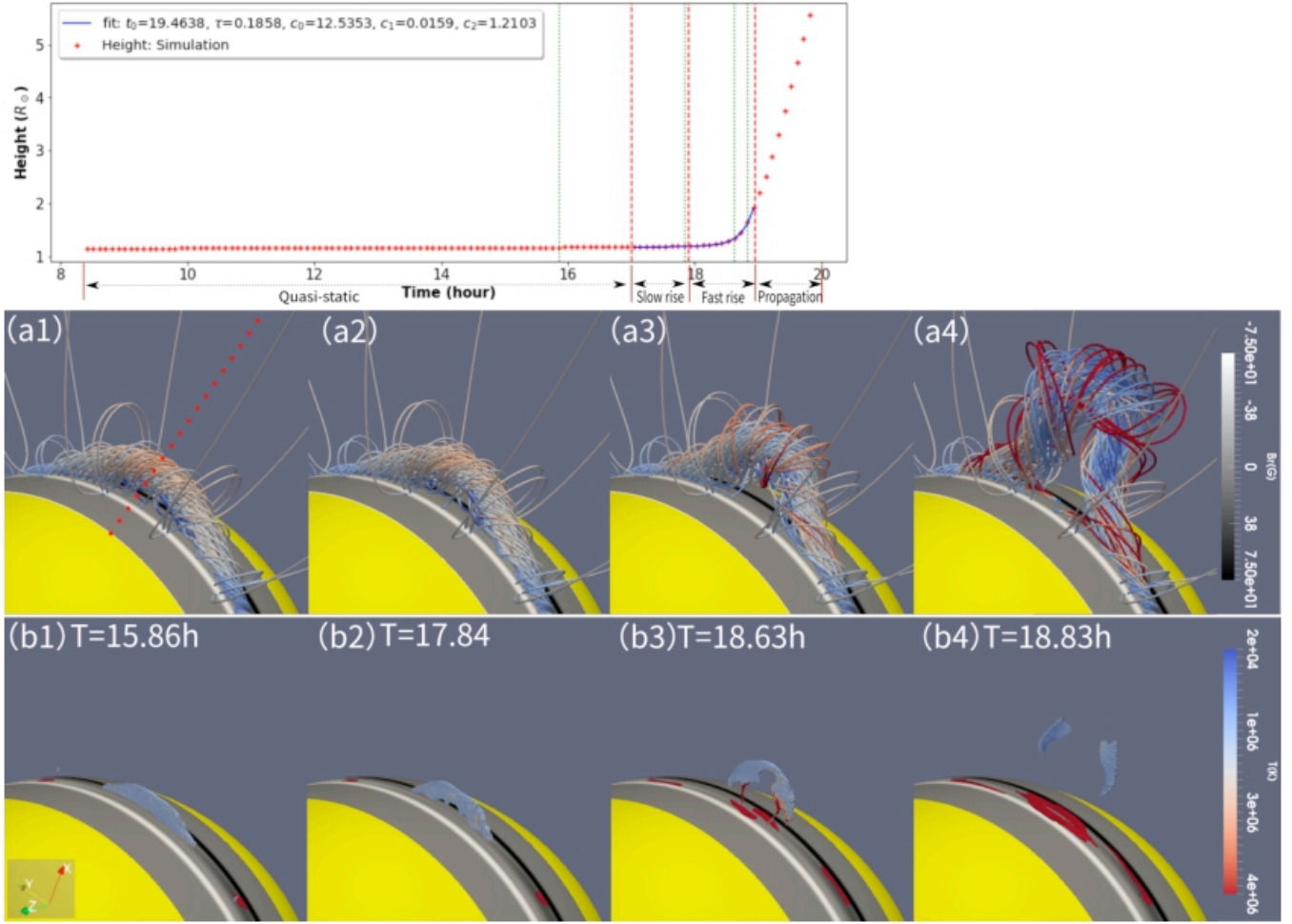

Figure 1. The top panel shows the height-time plot of the tracked apex of the axial field line of the emerged flux rope (marked as red plus sign) evaluated in Fan & Liu (2019). Slow rise and fast rise phases are covered by the fitting curve (in blue) which divides the evolution into the quasi-static (8.42 − 17 hour), slow rise (17 − 17.91 hour), fast rise (17.91 − 18.95 hour) and propagation (after 18.95 hour) phases separated by the three red dashed lines. The four green dotted lines mark the times of images in Panels (a1)-(a4). These field lines with fixed footpoints on the bottom (showing radial magnetic field) are colored by temperature. The corresponding density ($> 4 \times 10^{-15}$g/cm$^3$) are displayed in panels (b1)-(b4), which are also colored by temperature. The read dashed line marks the position of the middle vertical slice in Figure 2.

The J/B shown in panels c1-c3 of Figure 2 is a measure of the thickness of the current layer that develops. High J/B values are identified at the apex and dip regions with high Q values, which suggests that intense current layer tends to develop in high-Q regions. At the slow rise phase, an HFT at the apex region is identified as shown in the second column of Figure 2. The temperature in this region is higher than the surroundings, which is a signature of magnetic reconnection. The flux rope enters the fast rise phase and erupts once its axial field lines reaches the height of about 1.22 $R_\odot$ (inside the torus instability (Kliem & Török 2006) domain in Figure 4 of Fan & Liu (2019)) at about t = 17.91 hour. Soon after that, another HFT forms at the dip region under the flux rope (panel d3) which also shows high values of J/B (panel c3) indicating formation of intense current layers. The significant increase of temperature at the dip region HFT (panels a3, c3, d3) suggests the onset of explosive magnetic reconnection.

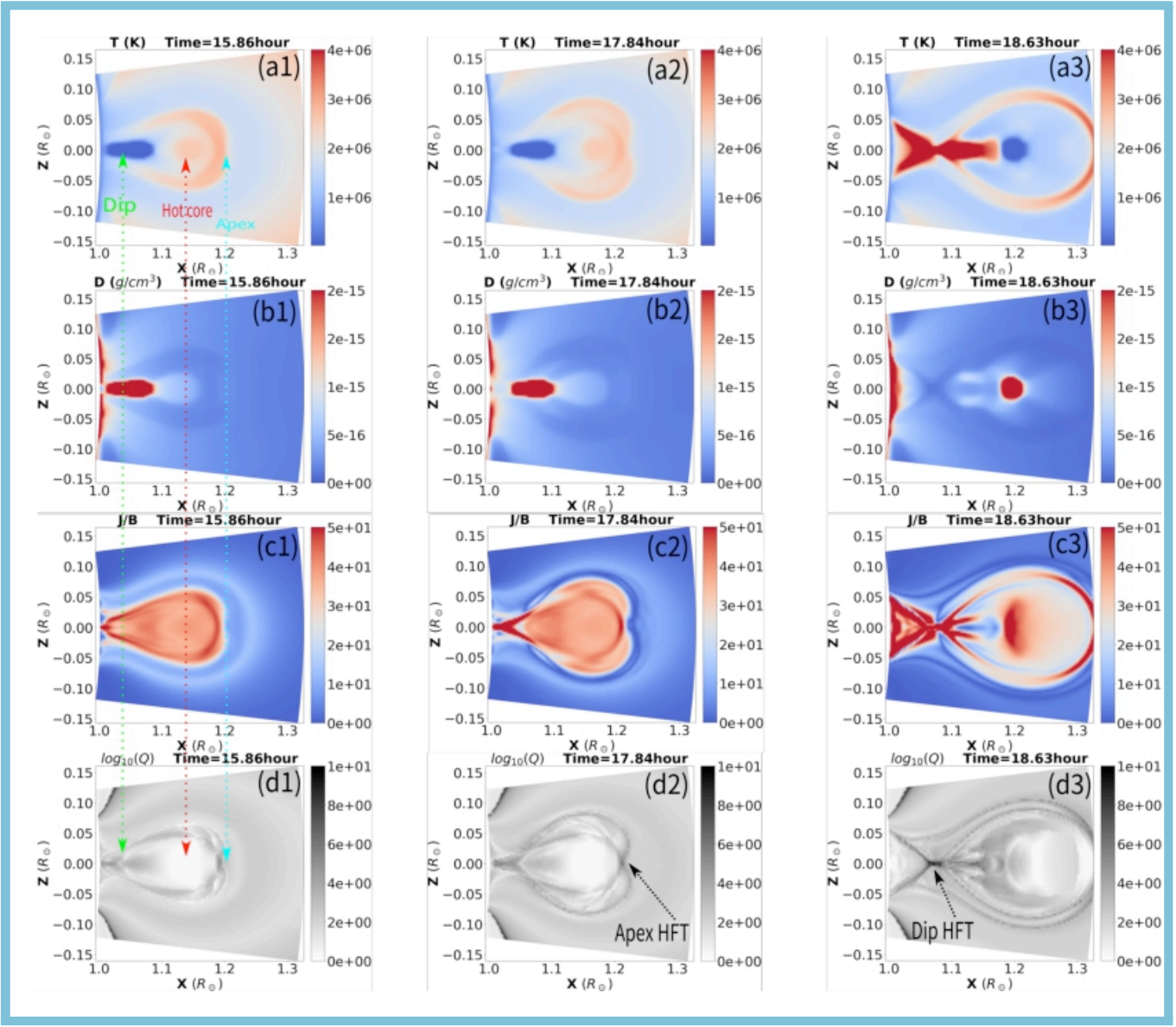

Figure 2. The distributions of temperature, density, J/B and Q factor on the middle vertical slice of the MFR (the first three panels of Figure 1) are displayed. The unit of the coordinate axes is solar radius. The dip, hot core and apex regions are marked by green, red and blue dotted lines with two arrows. The apex and dip HFTs are marked by black arrows in panels d2 and d3. An animation of the distributions of temperature, density, J/B and Q factor on the middle vertical slice of the MFR with the same field of view of panels (a1)-(d1), covering the time interval from t = 13.88 hour to t = 19.82 hour, is available. The real-time duration of the animation is 5 s.

Figure 3 presents the radial velocity ($V_r$) and forces ($F_{lorentz}$, $F_{hydro}$ and $F_{res} = F_{lorentz} + F_{hydro}$) along the central vertical line of the middle slice of the MFR (Figure 2) at t = 17.84 hour. The $F_{lorentz}$ is the radial Lorentz force of the magnetic field (i.e., the sum of the radial magnetic tension and magnetic pressure forces), and the $F_{hydro}$ refers to the sum of the gravity force and gas pressure gradient force. The locations with maximum (positive or negative) values of $dV_r/dr$ are determined as the locations where the flare reconnection (in the dip region) and overlying reconnection (in the apex region) occur, marked by two vertical red dotted lines in Figure 3. Diverging flows are detected in the dip region, which is represented by the negative (flow directing downward) and positive (flow directing upward) velocity values below and above the reconnection location, respectively. On the contrary, converging flows are detected in the apex region. The diverging flows and converging flows correspond to the outflows of the flare reconnection and the inflows of the overlying reconnection. A comparison of the top and bottom panels show that the bi-directional resultant force ($F_{res}$) are roughly consistent with the flows around the two reconnection locations. The diverging and converging flows are predominantly driven by the $F_{lorentz}$, while the $F_{hydro}$ is acting oppositely as shown in the middle panel of Figure 3. As mentioned above, signatures of magnetic reconnection in this simulation include localized heating at the apex and dip regions as well as bi-directional reconnection flows. The reconnection flows are very complex in three dimension, and will be investigated in detail in a future paper.

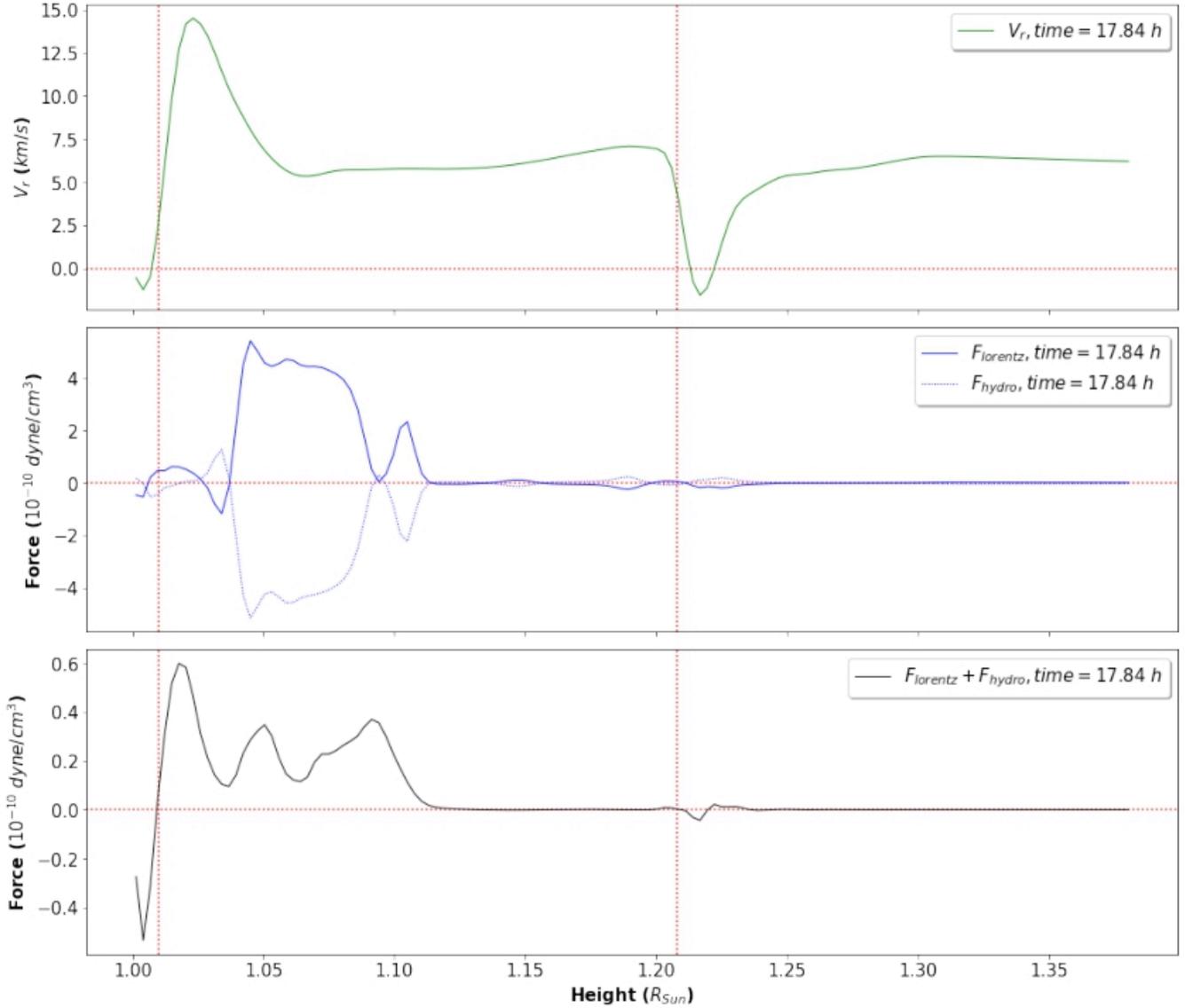

Figure 3. The radial velocity $V_r$, r-component of the Lorentz force $F_{lorentz}$ and hydrodynamic force $F_{hydro}$, as well as the resultant force $F_{res}$ = $F_{lorentz}$ + $F_{hydro}$ along the central vertical line of the middle slice of the MFR in Figure 2. The locations of the flare reconnection and overlying reconnection at t = 17.84 hour are marked by the two vertical red doted lines (Height = 1.010 and 1.208 R ).

At the slow rise phase (top panel of Figure 4), the maximum value (positive or negative) of the $F_{lorentz}$+$F_{hydro}$ at the apex region increases gradually, which indicates the enhancement of the overlying magnetic reconnection. At t = 17.84 hour, the location of the overlying magnetic reconnection reaches the height of 1.208 R . Once the eruption enters the fast rise phase, the $F_{lorentz}$ + $F_{hydro}$ (in the apex region) begins to decrease, while more prominent bi-directional $F_{res}$ is detected and sharply increases in the dip region (bottom panel of Figure 4). The sharply increasing $F_{res}$ is dominated by the magnetic tension force ($F_{ten}$) of the newly formed magnetic field lines due to the flare reconnection. These newly formed magnetic field lines possess very large curvature and opposite bending directions.

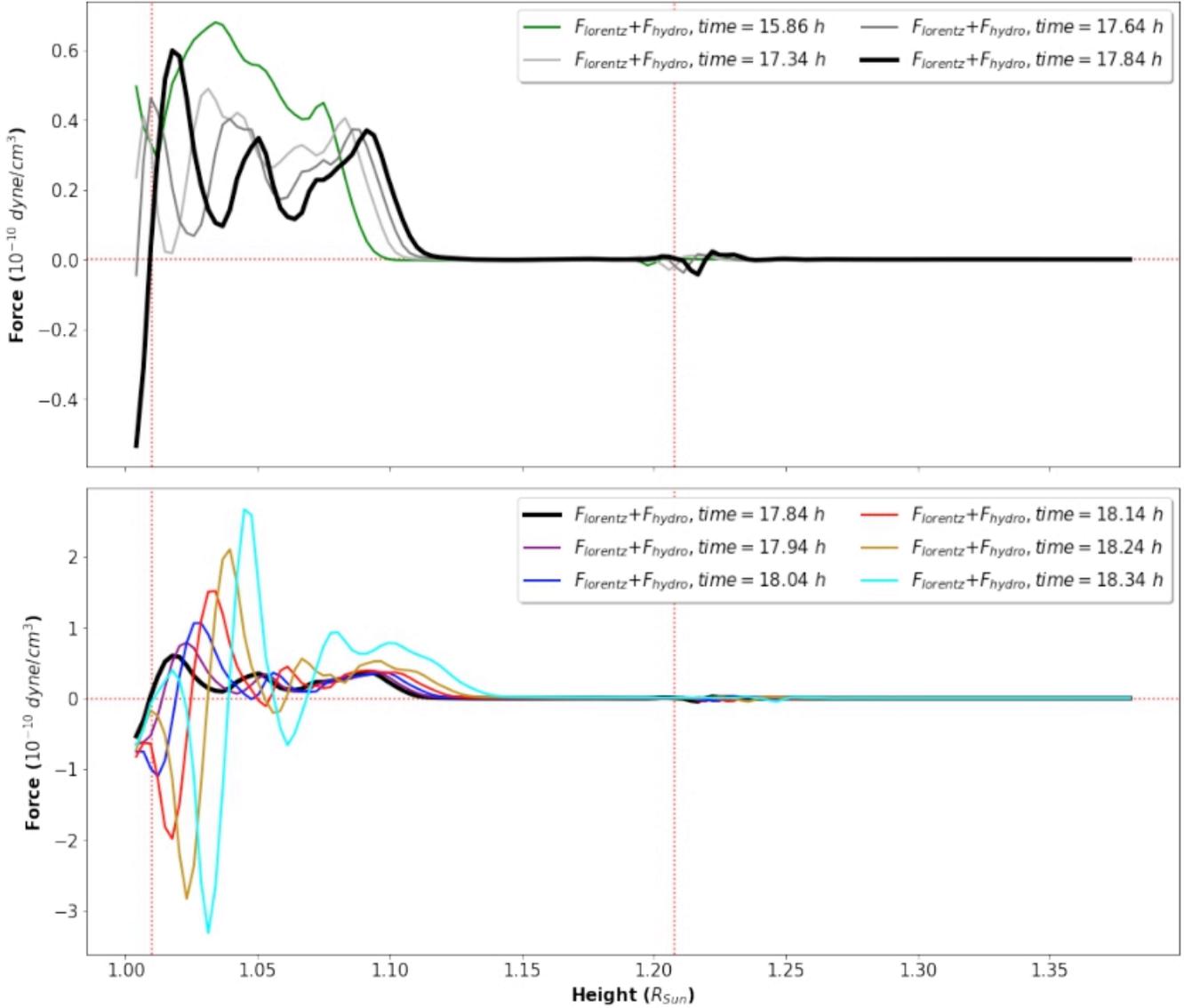

Figure 4. The evolution of the resultant force $F_{res} = F_{lorentz} + F_{hydro}$ at different time instances. The force at t = 17.84 hour marked as bold black lines serves as the boundary between the slow rise and fast rise phases. The two vertical red doted lines are the same as those in Figure 3. The top panel shows the $F_{res}$ at the quasi-static and slow rise phases, and the evolution of $F_{res}$ at the fast rise phase is displayed in the bottom panel.

The magnetic field is close to force-free during the quasi-static and slow rise phases when the $F_{lorentz}$ and $F_{hydro}$ balance with each other. The $F_{res}$ is too weak to raise the MFR which is confined by the overlying magnetic field. At the slow rise phase, the whole system can still be regarded as a nearly equilibrium state without obviously holistic resultant force and acceleration due to the confinement of the overlying magnetic filed. The term $F_{ten}^{int} = \int F_{ten} \times dl$ is proposed in order to measure the overlying confinement. The integral path is from the location of the overlying reconnection to the upper boundary of the simulation domain. The decrease of the $F_{ten}^{int}$ during the slow rise phase (left panel of Figure 5) is due to the removal of the overlying confinement by overlying reconnection. The nature of the slow rise is the quasi-static evolution of the MFR whose upper boundary slowly expands and merges with the overlying magnetic field due to overlying reconnection occurring between the upper boundary of the MFR and the overlying magnetic field. During the fast rise phase, the right panel of Figure 5 shows that the resultant force $F_{res}^{int}$ integrated along the central vertical line (from the location of the flare reconnection to the upper boundary of the simulation domain) is upward and rapidly growing, which drives the fast rise of the MFR.

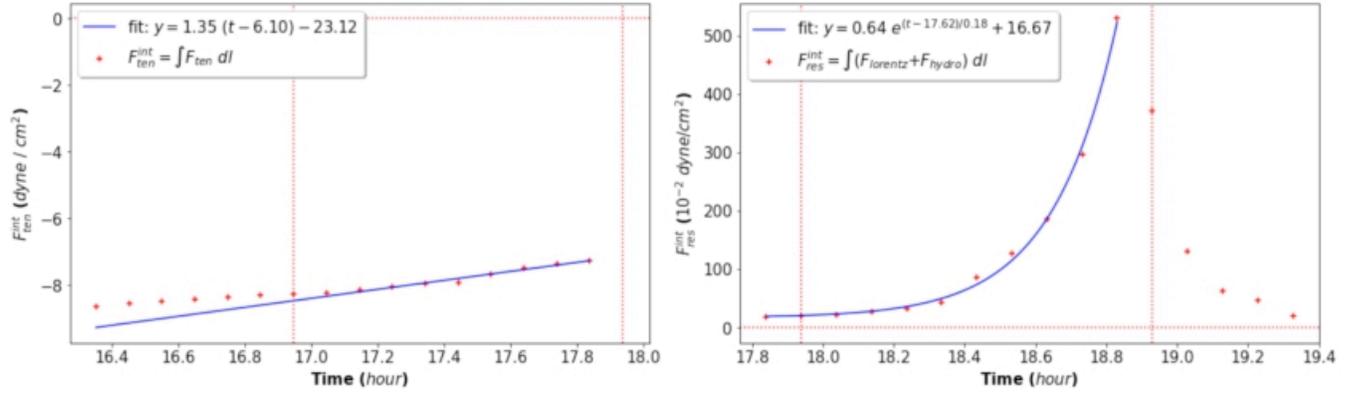

Figure 5. The evolution of the $F_{ten}^{int}$ during the slow rise phase and the $F_{res}^{int}$ during the fast rise phase. The values of $F_{ten}^{int}$ and $F_{res}^{int}$ in this simulation are marked by the red '+', which are fitted by the linear and exponential functions represented by the blue line and curve. The corresponding fitting functions are listed at the upper left corners. The vertical red dotted lines mark the boundaries of slow rise and fast rise phases similar to Figure 1.

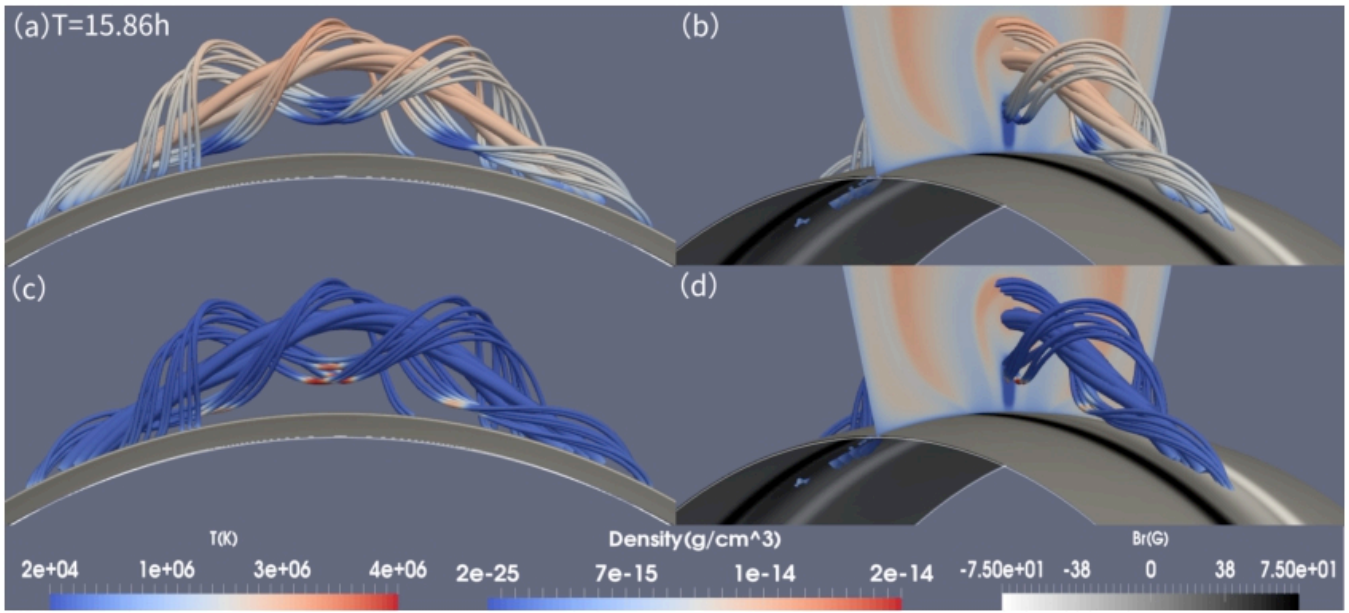

Figure 6. Panels a and c show field lines going through the filaments (dips), hot core and hot cavity (apex) colored by temperature and density, respectively. The bottoms of panels a-d show radial magnetic field. Temperature is displayed on the vertical slice.

In summary, we provide a semi-quantitative description to explain why the eruption can be divided into the slow rise and fast rise phases, and what are the main driving mechanisms for these two phases. The slow rise of the MFR is driven by the linear decrease of the $F_{ten}^{int}$, which indicates that the overlying reconnection occurs with a nearly constant velocity. During the quasi-static phase, the decrease of the $F_{ten}^{int}$ is much slower, which is likely due to the much slower overlying reconnection. The fast rise is driven by the rapidly increasing $F_{res}^{int}$ from the newly formed magnetic field lines due to the explosive flare reconnection, which drives an exponential eruption of the MFR. After the fast rise phase, the $F_{res}^{int}$ rapidly decreases and the eruption enters the propagation phase. In the following section, we examine the two types of magnetic reconnections occurring in the apex and dip regions, respectively.

## 5. THE 3D QSLS AND MAGNETIC RECONNECTION

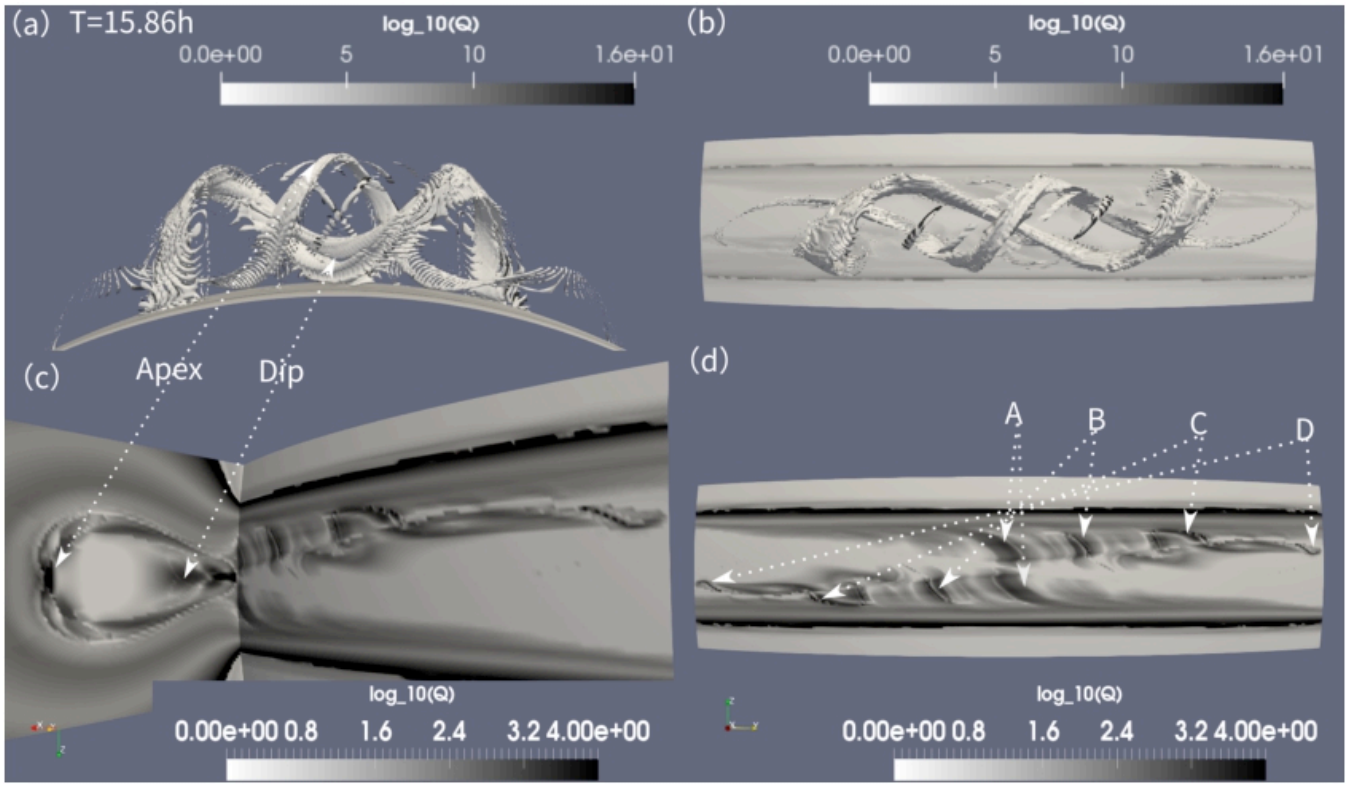

Figure 7. Two 3D isosurfaces ($\log_{10} Q = 3$) of high Q factor look like two ribbons marked as 'Apex' and 'Dip' in panels a and b. Panels c and d show the corresponding vertical and bottom Q factor slices. 4 pairs of high Q regions (A, B, C and D) are marked in the bottom.

Figure 6 shows three sets of magnetic field lines passing through the hot core, the apex region, and the dip region in the central vertical slice shown in the first column of Figure 2. The hot core region consists of twisted field lines without any dips, which pass through the center of the flux rope. The two sets of field lines passing through the apex and dip regions are both prominence-carrying, dipped field lines that spiral around the central core field lines and approach the boundary of the MFR. The prominence materials are located at the dips where the temperature is low, while the apex of the same field lines possesses low density and high temperature. Figure 7 shows the isosurface of the computed squashing factor Q (with $\log_{10} Q = 3$), which outlines the 3D structure of the regions with high Q values, i.e., the QSLs. We find two ribbons of high-Q regions (panels a and b), one of which passes through the dip region in the central vertical slice (panel c), which we call the "dip ribbon", and the other passes through the apex region in the central slice (panel c), which we call the "apex ribbon". The dip ribbon has one dip part and two apex parts, while the apex ribbon possesses two dip parts and three apex parts. The two high-Q ribbons roughly coincide with the two sets of prominence-carrying, spiraling boundary field lines shown in Figure 6. Panel d of Figure 7 shows four pairs of high Q regions at the bottom boundary surface, marked as regions A, B, C and D. The apex ribbon touches the bottom at regions B and D, while the dip ribbon touches the bottom at regions A and C. Figure 8 shows that the two high-Q ribbons go through the hot cavity boundary. The regions with the highest temperature in the cavity coincide with the apex parts of the two high-Q ribbons, while the prominence condensation is located at the dips of the high-Q ribbons. The high temperature in the cavity boundary implies the occurrence of continuous magnetic reconnections. The central hot-core field lines also show relatively high temperature, although not as high as the cavity boundary. However, the Q value near the hot-core is low (see also panel d1 in Figure 2). The high temperature of the hot core is likely due to the numerical dissipation of the distributed volume current along the twisted core field lines.

Magnetic reconnection preferentially occurs at the high-Q regions or HFTs (e.g. Savcheva et al. 2012; Liu et al. 2018) such as the central apex and dip regions of the QSL ribbons. To investigate the magnetic reconnection taking place at the HFTs, we show two examples. One of which takes place at the central apex region of the apex ribbon, and the other occurs at the central dip region of the dip ribbon. In the central vertical slice of the system shown in Figure 9, we draw two magnetic field lines along the apex ribbon passing through the apex region, with their left footpoints fixed at the bottom. The short and sheared arcade-like field line (coming from the overlying arcades) with two footpoints located in region B is located slightly above the long and helical field line (coming from the flux rope) with two footpoints located in region D. We find that the right footpoint of the long field line jumps from region D to B (panels a2 and b2) from t = 17.34 to 17.64 hour, while the right footpoint of the short field line moves from region B to D (panels a2 and b2) for the same time interval, during the slow rise phase. The short field line obtains a dip while the long field line loses a dip. The motion of the footpoints of the field lines and the change of the field line morphology are the results of magnetic reconnection. These overlying MRs between the flux rope and the overlying arcades at the apex HFT are similar to the breakout reconnections at the overlying null point in a multipolar configuration described by Antiochos et al. (1999), since both of the reconnections play a role in removing the confinement of the overlying field. However, the magnetic configuration in our simulation includes a narrow bipolar band with an emerging flux rope, and the overlying reconnections occurred at the apex HFT, while the breakout reconnection occurred at the overlying null point in a quadrupolar configuration.

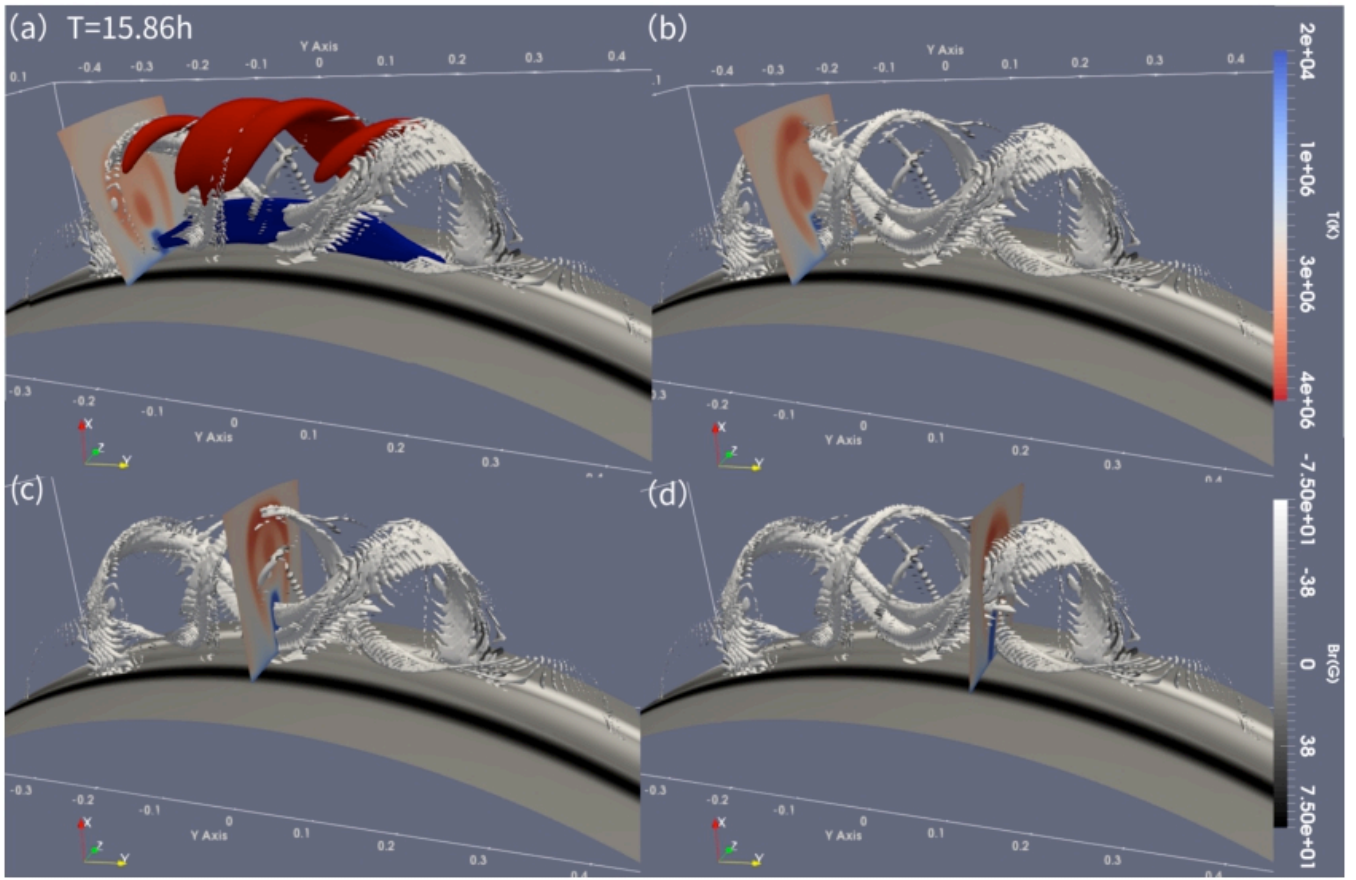

Figure 8. Same as Figure 7 but from a different view. A vertical slice colored by temperature moves along the two 3D high Q ribbons shown in panels a-d. Panel a shows the corresponding temperature isosurfaces colored by red (T = 2600000K) and blue (T = 75000K) which represent the heating region and the filament, respectively.

In Figure 10, we show another two field lines passing through the HFT at the central dip region during the fast rise phase. The left footpoints of the two field lines are fixed. In panels a1-a2 and b1-b2, we find that the right footpoint of the left field line jumps from regions A, B to region C, while the right footpoint of the right field line jumps from region C to between regions A, B. Panels a3 and b3 show that the two field lines are close to each other before the reconnection, and then separate from each other after the reconnection. The two J-shape field lines without dips translate to one long sigmoidal field line with a dip and a short low-lying loop field line (panels a1-a2 and b1-b2), which is consistent with the classical tether-cutting reconnection proposed by Moore et al. (2001).

In summary, we identify two kinds of MRs which occur at the apex and dip parts of the two high-Q ribbons, respectively. During the slow rise phase, overlying MRs occur at the apex parts of the QSL ribbons, which removes the confinement of the overlying magnetic field. The fast rise of the flux rope begins once its apex enters the torus instability domain. The tether-cutting flare reconnection occurs at the dip region, which leads to the change of the connectivity of magnetic field lines and contributes to the increase of twist in the MFR. The direction of the magnetic tension force above the reconnection location is upward after reconnection, which enhances the acceleration of the MFR and explosive release of magnetic energy.

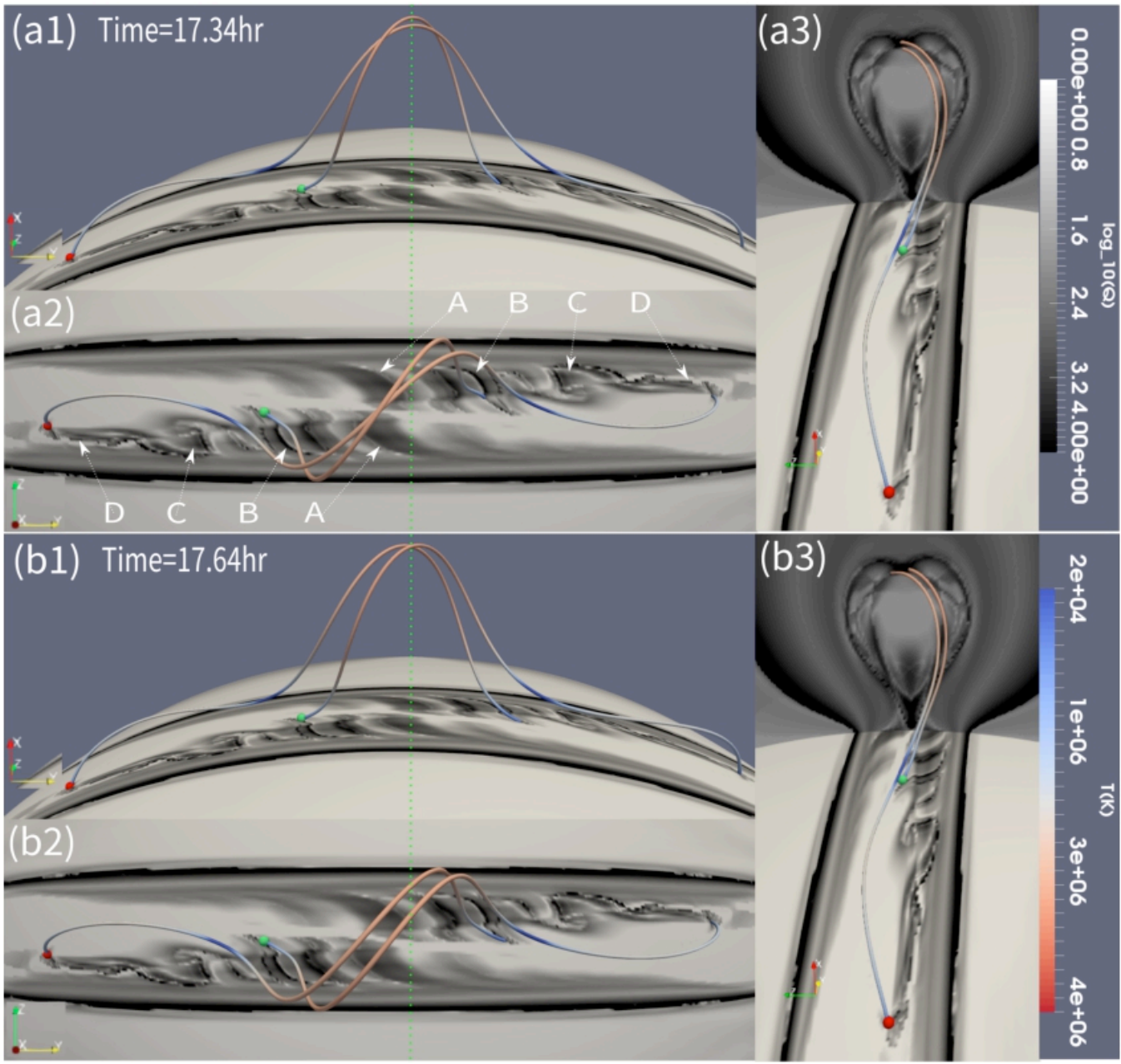

Figure 9. Panels a1-a3 (before reconnection) and b1-b3 (after reconnection) show two magnetic field lines colored by temperature, going through the apex region of the cavity on the middle vertical slice, with their left footpoints (marked by red and green balls) fixed on the bottom, during the slow rise phase. The distribution of Q factor is displayed on the bottom and vertical slices. The dashed green line marks the position of the vertical slice whose left view is displayed in panels a3 and b3. High Q regions on the bottom are marked as A, B, C and D in panel a2.

Figure 11 shows the evolution of three magnetic field lines with fixed left foot points which originally pass through the cold and dense filaments at different heights on the vertical slice. The highest field line rises into the hot core and then continues to rise toward the apex of the hot cavity (panels b and c). The middle field line transits from dipped prominence carrying field line into a field line without prominence in the cavity as it loses its dip. The lowest field line remains as a prominence carrying field line but rises from the middle of the prominence to the top (panels a-c). The three field lines all lose dips during this evolution (panels d-f). This evolution indicates that some of the prominence-carrying magnetic field lines transit to "horn" field lines and then to hot core field lines as they lose their dips. This apparent relative movement between the prominence and the magnetic field has also been identified in observations (e.g., Cheng et al. 2014). The underlying cause is likely that the prominence condensations evaporate to higher temperature and expand along the field lines, when the dips disappear as the flux rope rises.

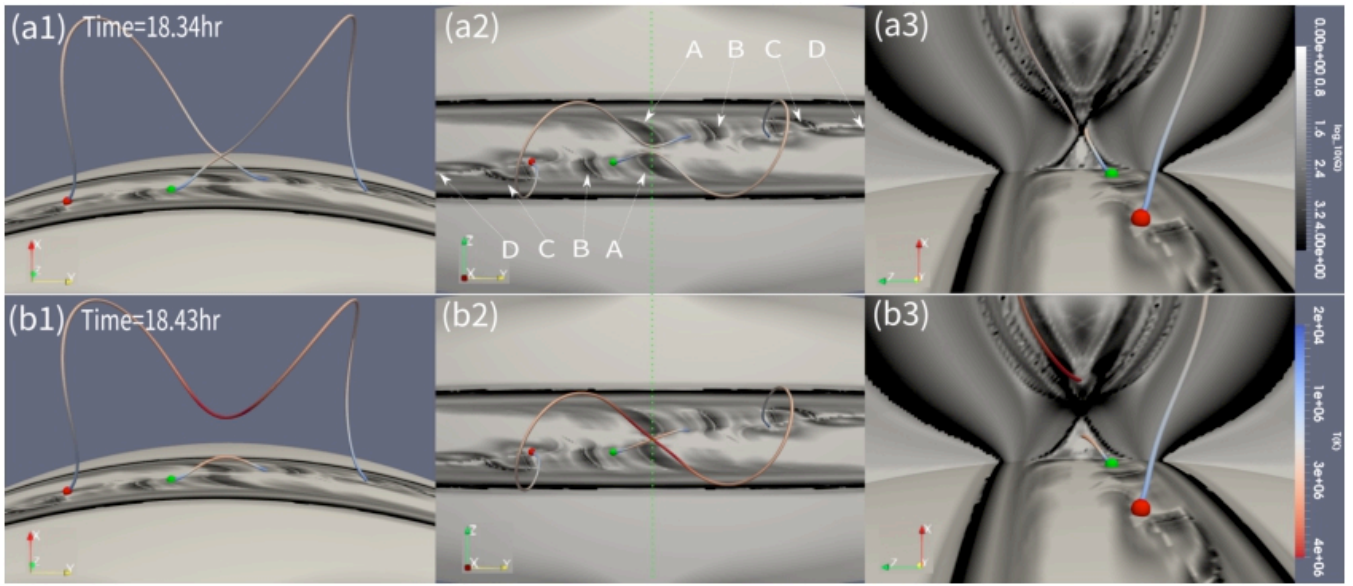

Figure 10. Same as Figure 9 but for the magnetic field lines going through the dip region of the cavity on the vertical slice during the fast rise phase.

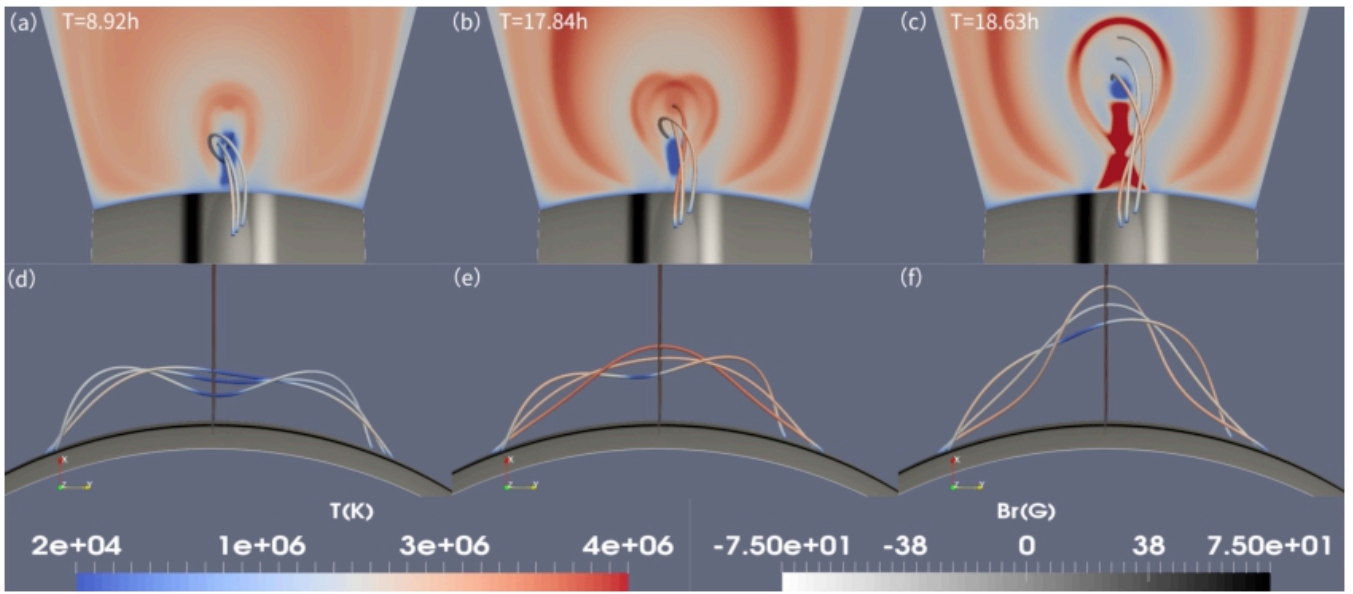

Figure 11. Three magnetic field lines colored by temperature going through the filaments with left footpoints fixed are displayed. The middle vertical slice shows the temperature and the radial magnetic field is displayed on the bottom in panels a-c. Another view is displayed in panels d-f.

6. SUMMARY AND DISCUSSIONS

In this paper, we study the magnetic reconnection in a 3D MHD simulation of the prominence-cavity system in which various observed features such as prominence "horns" and hot cores in the coronal cavity are identified. At the lower boundary, a pair of bipolar bands of magnetic flux are employed to construct a coronal streamer, under which a twisted MFR is built up quasi-statically by imposing the emergence of a twisted magnetic torus at the bottom boundary. The filaments are located at the dip region of the MFR. The hot core is located above the cold and dense filaments and within the cavity. As discussed in Fan & Liu (2019), the prominence-cavity system in this simulation is more in accordance with the observed quiescent prominence-cavity systems. According to the height-time plot of the flux rope, the evolution of the prominence-cavity system can be divided into four phases: quasi-static phase (8.42 − 17 hour), slow rise phase (17 − 17.91 hour), fast rise phase (17.91 − 18.95 hour) and propagation phase (after 18.95 hour). This division is consistent with the observations of solar eruptions such as Zhang et al. (2001) and Cheng et al. (2020). The role of various mechanisms (ideal and resistive) at each phase is still a matter for debate, which may depend on both the magnetic configuration being studied and the numerical model describing the temporal evolution. Moreover, the dominant mechanism may vary from event to event. We explore the dominant mechanisms during different phases of the eruption in this simulation. A semi-quantitative description is provided to explain why the eruption can be divided into the slow rise and fast rise phases and what are the major driving mechanisms for these two phases.

Magnetic topology analysis of the prominence-cavity system shows that the squashing factor Q is the highest at the apex and dip HFTs which extend to the apex and dip ribbons in 3D. Therefore, magnetic reconnection mainly occurs along the two high-Q ribbons, which is confirmed by the corresponding temperature increase of the hot cavity boundary and dip regions. Two kinds of magnetic reconnections are identified and occur along the two high-Q ribbons (dip ribbon and apex ribbon). The total twist of the emerged MFR is about 1.76 winds after the flux emergence is stopped at t = 8.42 hour while the slow rise of the MFR begins at t = 17 hour when the overlying reconnections become apparent. Kink instability (Kliem et al. 2004) does not occur at or engine the slow rise, because the prominence weight suppresses the development of kink instability (Fan 2020). During the slow rise phase, the overlying reconnection between the MFR and the overlying arcade occurs at the apex HFT. The integrated magnetic tension force (above the location of the overlying reconnection) linearly decreases with the overlying reconnection, which removes the overlying confinement and contributes to the slow rise of the MFR. The nature of the slow rise is the quasi-static evolution of the MFR whose upper boundary slowly expands and merges with the overlying magnetic field lines. The flux rope enters the explosive fast rise phase, once the torus instability is satisfied and the flare reconnection occurs at the HFT in the dip region. The integrated resultant force (above the location of the flare reconnection) from the newly formed magnetic field lines due to the flare reconnection exponentially increases, which drives the fast rise of the MFR.

Hassanin & Kliem (2016) studies confined solar eruptions in a zero β MHD simulation, and finds two distinct phases of strong magnetic reconnections which first occurs in the helical current sheet between the twisted MFR and the overlying magnetic field, and then between the legs of the MFR in the vertical current sheet. Similarly, we also find two distinct phases in the simulation of a erupting prominence-cavity system, i.e., the dominant overlying reconnection at the apex HFT during the slow rise phase and the dominant flare reconnection at the dip HFT during the fast rise phase. The difference is that the prominence-cavity system in our simulation contains the formation of prominence condensations which have a significant effect on the stability of the MFR (Fan 2020). Another difference is that the MFR successfully erupts in our simulation. The two studies suggest that different magnetic reconnections occurring in different phases exists in both eruptive and confined eruptions, and whether torus instability is satisfied or not, will leads to a successful eruption or a confined eruption.

MHD instabilities especially torus instability are often identified as the main initiation mechanism for the onset of the fast rise phase. Through a zero β MHD simulation of a successful eruption, Aulanier et al. (2010) finds that photospheric flux-cancellation and tether-cutting coronal reconnection play a dominant role in the build up and rise of the pre-eruptive flux rope, while the torus instability is the main cause for the eruption. Similar conclusions have also been drawn by Cheng et al. (2020) who carries out a thorough investigation on the initiation and early kinematic evolution of 12 solar eruptions including both hot channels and quiescent filament eruptions. A recent statistical study by Zou et al. (2019) suggests that filaments on the quiet Sun as well as intermediate filaments are more likely to be triggered by an ideal MHD process, due to the lack of reconnection features. ? has demonstrated that the catastrophe and the torus instability are equivalent descriptions for the onset condition of solar eruptions. Through a 2.5D model analysis, ? demonstrates that ideal MHD catastrophe alone can produce CMEs, even fast ones. Nevertheless, the onset of magnetic reconnection can significantly enhance the eruptive speed. On the other hand, a study of 2.5D simulations of breakout CMEs in a quadrupolar magnetic configuration by ? shows that the fast reconnections at the breakout and flare current sheets lead to the CME onset and the explosive CME acceleration, respectively.

In our simulation, the overlying reconnection with a nearly constant velocity at the apex HFT plays an important role in bringing the system to the regime of torus instability, and the flare reconnection quickly switches on at the dip HFT which further enhances the explosive fast rise. This process likely plays an important role in the quiescent prominence eruptions which do not meet the conditions of kink instability and show no signatures of tether-cutting reconnection before eruption. In fact, the main process in this simulation is consistent with that identified in an observed polar crown prominence eruption (Su et al. 2015), during which the prominence supported by a flux rope enters the regime of torus instability at the onset of the fast-rise phase, and signatures of magnetic reconnection below the prominence appear about one hour later. To validate the role of the overlying reconnection during the slow rise phase and the flare reconnection during the fast rise phase, further detailed observational investigations and data-constrained or data-driven MHD simulations are required.

## 7. ACKNOWLEDGEMENTS

We thank Dr. Yuhong Fan for providing the simulation model data of Fan & Liu (2019) and discussing the results of this paper extensively. We would like to thank the anonymous referee for his/her careful reading and valuable comments, which help to improve this manuscript. We thank Drs. Yang Guo and Xin Cheng for helpful discussions about magnetic topology analysis and curve fitting. Graduate student Tie Liu is supported by the scholarship from the Chinese Scholarship Council of the Ministry of Education of China. This work is also supported by the National Natural Science Foundation of China grant NO. 41761134088, 11790302 (11790300), 11473071, and U1731241,as well as the Strategic Priority Research Program on Space Science, CAS, Grant No. XDA15052200 and XDA15320301.